\documentclass[conference]{IEEEtran}
\IEEEoverridecommandlockouts
\usepackage{cite}
\usepackage{amsmath,amssymb,amsfonts}
\usepackage{algorithmic}
\usepackage{graphicx}
\usepackage{textcomp}
\usepackage{xcolor}
\def\BibTeX{{\rm B\kern-.05em{\sc i\kern-.025em b}\kern-.08em
    T\kern-.1667em\lower.7ex\hbox{E}\kern-.125emX}}
\begin{document}

\title{Realtime Person Identification via Gait Analysis}

\makeatletter
\newcommand{\linebreakand}{%
  \end{@IEEEauthorhalign}
  \hfill\mbox{}\par
  \mbox{}\hfill\begin{@IEEEauthorhalign}
}
\makeatother

\author{\IEEEauthorblockN{Shanmuga Venkatachalam}
\IEEEauthorblockA{\textit{ECE Department} \\
\textit{Carnegie Mellon University}\\
shanmugv@andrew.cmu.edu}
\and
\IEEEauthorblockN{Harideep Nair}
\IEEEauthorblockA{\textit{ECE Department} \\
\textit{Carnegie Mellon University}\\
hpnair@andrew.cmu.edu}
\and
\IEEEauthorblockN{Prabhu Vellaisamy}
\IEEEauthorblockA{\textit{ECE Department} \\
\textit{Carnegie Mellon University}\\
pvellais@andrew.cmu.edu}
\and
\IEEEauthorblockN{Yongqi Zhou}
\IEEEauthorblockA{\textit{ECE Department} \\
\textit{Carnegie Mellon University}\\
yongqiz2@andrew.cmu.edu}
 \linebreakand
\IEEEauthorblockN{Ziad Youssfi}
\IEEEauthorblockA{\textit{ECE Department} \\
\textit{Carnegie Mellon University}\\
zyoussfi@andrew.cmu.edu}
\and
\IEEEauthorblockN{John Paul Shen}
\IEEEauthorblockA{\textit{ECE Department} \\
\textit{Carnegie Mellon University}\\
jpshen@andrew.cmu.edu}
}

\maketitle

\begin{abstract}
Each person has a unique gait, i.e., walking style, that can be used as a biometric for personal identification. Recent works have demonstrated effective gait recognition using deep neural networks, however most of these works predominantly focus on classification accuracy rather than model efficiency. In order to perform gait recognition using wearable devices on the edge, it is imperative to develop highly efficient low-power models that can be deployed on to small form-factor devices such as microcontrollers. In this paper, we propose a small CNN model with 4 layers that is very amenable for edge AI deployment and realtime gait recognition. This model was trained on a public gait dataset with 20 classes augmented with data collected by the authors, aggregating to 24 classes in total. Our model achieves 96.7\% accuracy and consumes only 5KB RAM with an inferencing time of 70 ms and 125mW power, while running continuous inference on Arduino Nano 33 BLE Sense. We successfully demonstrated realtime identification of the authors with the model running on Arduino, thus underscoring the efficacy and providing a proof of feasiblity for deployment in practical systems in near future.
\end{abstract}

\begin{IEEEkeywords}
Human Gait, Biometric Identification, Inertial Sensors, Arduino, Neuromorphic Akida, Edge AI
\end{IEEEkeywords}

\section{Introduction}

In the field of biometric identification, traditional methods such as fingerprint, facial recognition dominate. However, gait analysis is fast emerging as a unique and promising approach for identifying a person. Gait, the distinctive way an individual walks, carries inherent characteristics that can be leveraged for accurate non-intrusive person identification \cite{sepas2022deep,Lynnerup2005PersonIB}.
%
Unlike static biometrics, such as fingerprints and facial features, gait analysis taps into the dynamic and behavioral aspects of an individual's movement. Every person has a distinct gait, influenced by factors like anatomy, musculoskeletal structure, and personal habits. This distinctiveness makes gait analysis an intriguing and effective tool for identifying individuals in diverse settings, ranging from surveillance and security to healthcare and rehabilitation.

For this course project (Figure \ref{framework}), we experimented on using light-weight convolutional neural network (CNN) models for edge-based gait detection for person identification. We perform pre-processing of the raw gait signals and model the CNN on the Edge Impulse framework. We use a popular gait dataset and further augment it with raw data collected from our team to train and test our model. We deploy our model on an Arduino Nano BLE 33 board for live inference and demonstration. We demonstrate highly accurate gait detection through our results and performed a live demonstration to show its efficacy. Further, we also deploy on smartphone.

Finally, we convert our CNN model to its event-based spiking neural network (SNN) equivalent via Brainchip MetaTF framework and deploy the SNN to the Brainchip Akida processor \cite{akidabc}. We obtain real-time power and latency measurements.

\begin{figure}[t]
    \centering
    \includegraphics[width=\columnwidth]{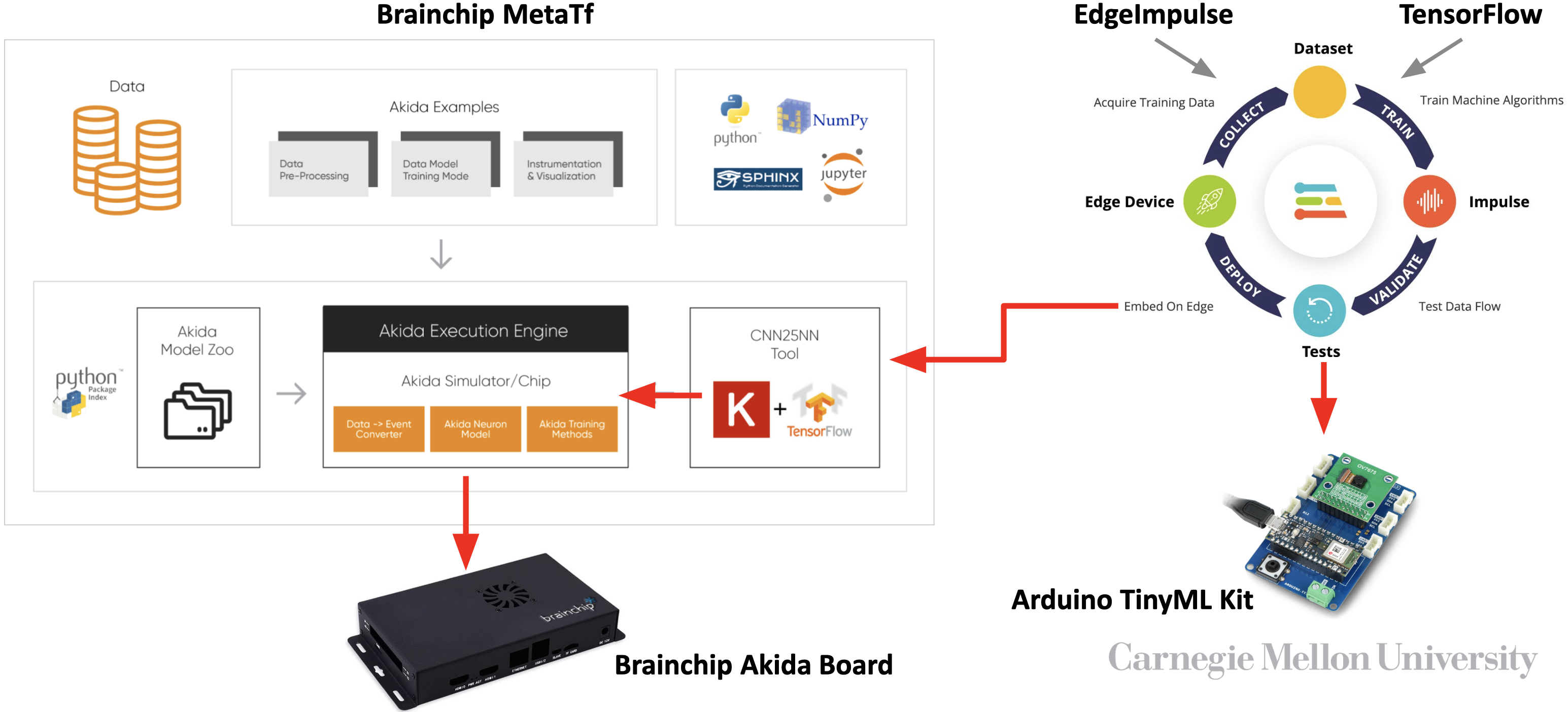}
    \caption{Overall Framework}
    \label{framework}
\end{figure}

\section{Methods}
This section details the methods and experimental setup undertaken for this work. First, the dataset including our custom data collection procedure is described, followed by the edge inferencing pipeline, and finally our training methodology.

\subsection{Dataset}
We use whuGAIT dataset \cite{zou2020deep}. A total of 118 subjects participated in the data collection process. Within this group, 20 subjects gathered data over a span of two days, generating thousands of samples each. Simultaneously, 98 subjects undertook a more concise data collection, spanning one day and resulting in hundreds of samples each. Each data sample comprises both 3-axis accelerometer and 3-axis gyroscope data, all recorded at a uniform sampling rate of 50 Hz. 

\begin{figure}[t]
    \centering
    \includegraphics[width=0.9\columnwidth]{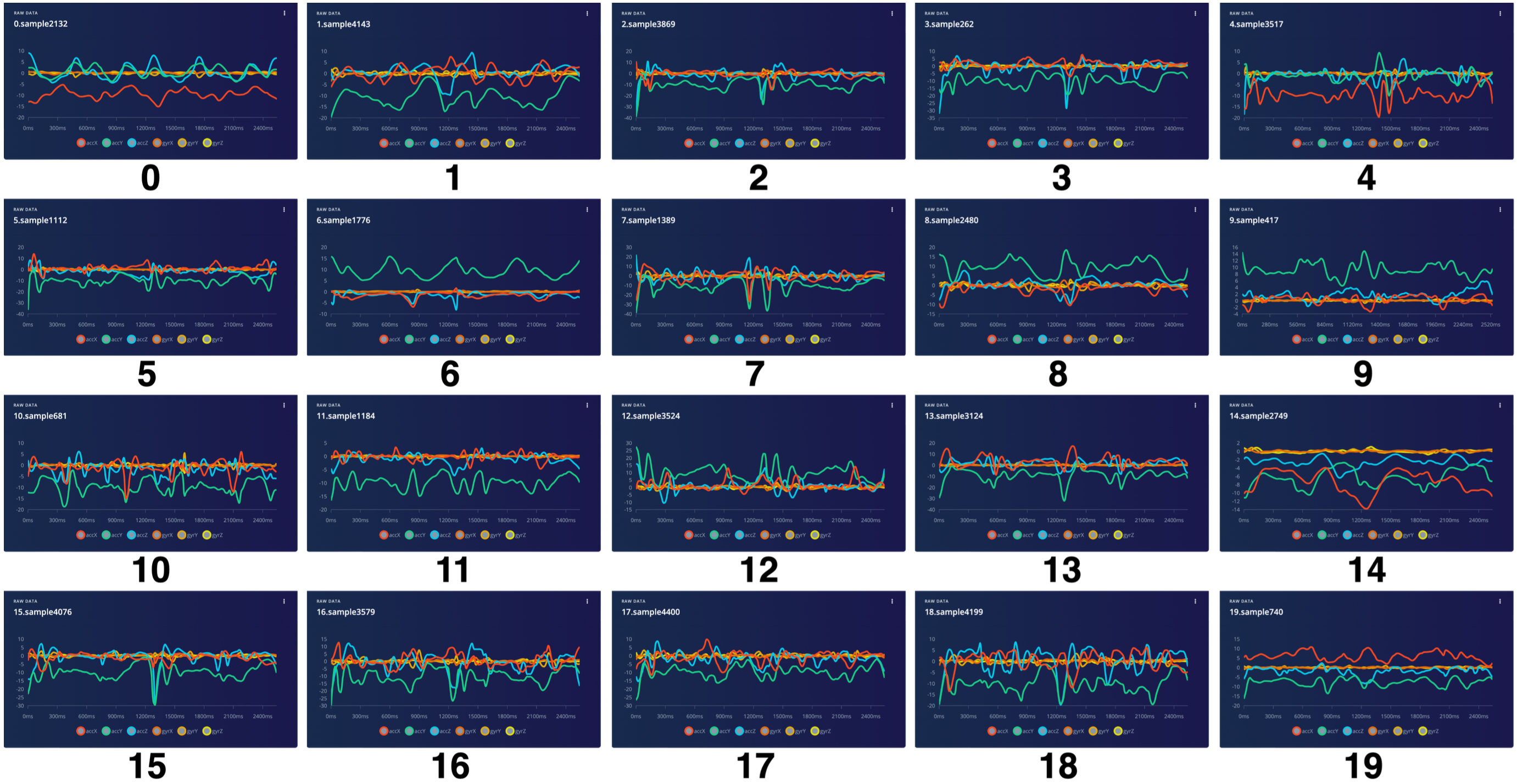}
    \caption{Raw data example from each of the 20 classes}
    \label{dataset}
\end{figure}

\begin{figure}[t]
    \centering
    \includegraphics[width=0.75\columnwidth]{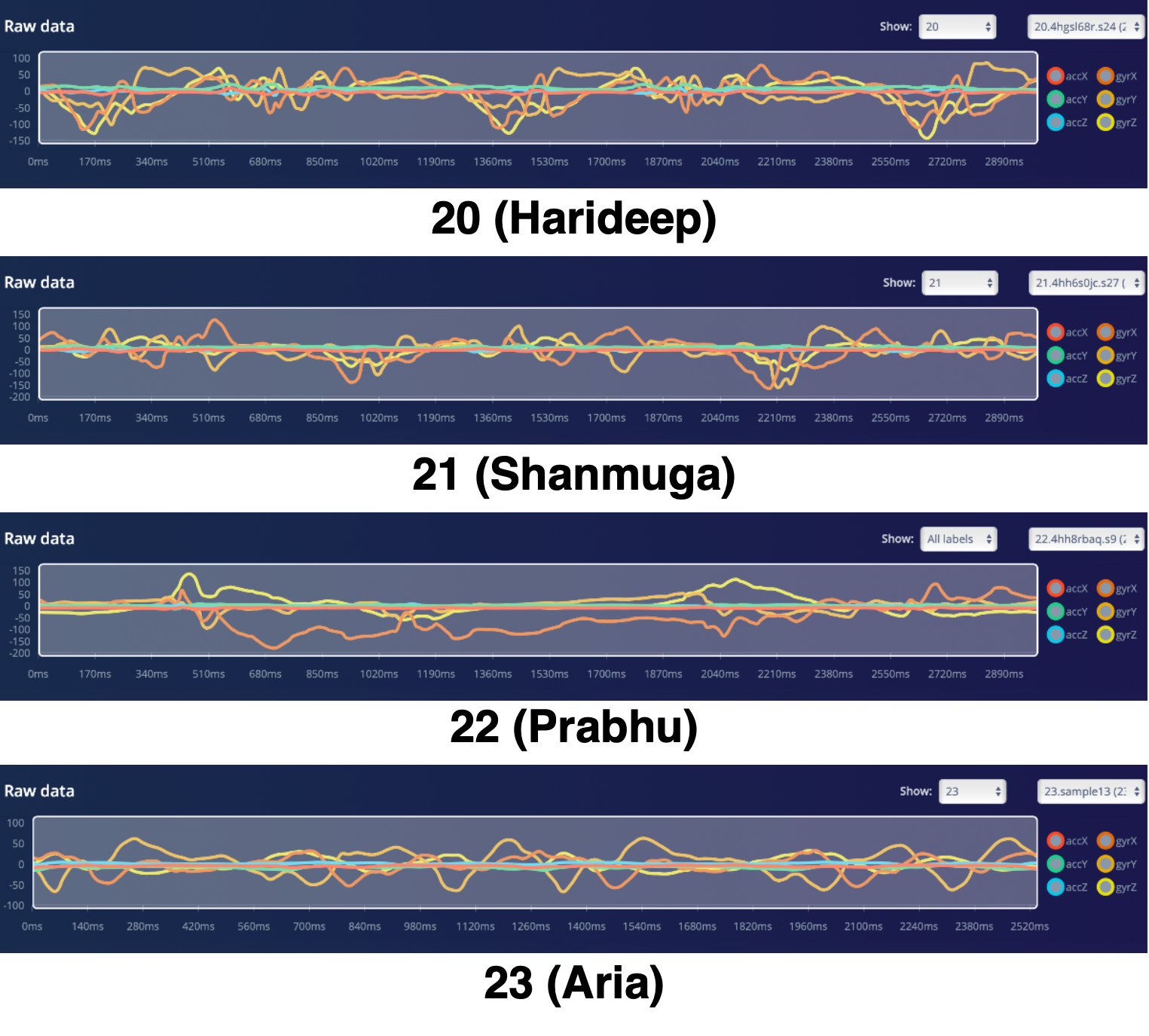}
    \caption{Raw data example from each of the 4 new augmented classes}
    \label{dataset_custom}
\end{figure}

Among the various sub-datasets, we focus on dataset\#2, since it consists of only 20 subjects as oppposed to 118 for simplicity, where the gait curve is divided into two-step samples and interpolated into length 128. It consists of 49,275 samples, of which 44,339 samples are used for training and the rest 4,936 for testing. One example of raw data from each class is shown in Figure \ref{dataset}.

Further, in addition to incorporating the public dataset, we collected data from each of the four authors, increasing the overall class count to 24. The four custom classes augmented to the dataset are exemplified in Figure \ref{dataset_custom}. It is to be noted that data for the first 20 classes have been collected from IMU sensors in smartphone whereas the custom data for the last four classes have been collected with IMU sensors in Arduino, where the arduino while still being connected to laptop was placed inside pant pocket while walking holding the laptop. The data was collected for walking at multiple paces back and forth, on carpeted as well as non-carpeted floors for better generalization. Our final model uses spectral feature extraction on this raw data for ease of live deployment and demonstration, however we have also created a custom data preprocessing pipeline as explained next.

Our initial implementation involves manually splitting the collected data into 3-second gait segments using the visualization tool provided by Edge Impulse. However, this is inconvenient for collecting large amounts of data since it requires manual data processing.

\begin{figure}[t]
    \centering
    \includegraphics[width=0.8\columnwidth]{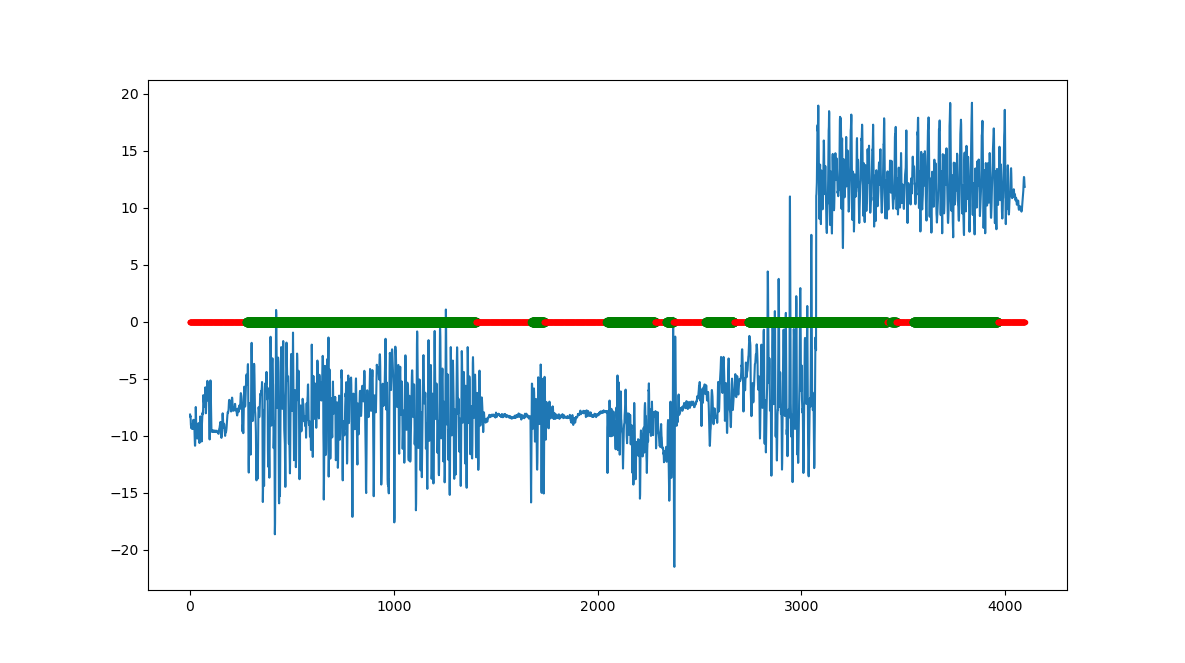}
    \caption{Mobile Phone Deployment: DCNN output}
    \label{fig:dcnnRes}
\end{figure}
\begin{figure}[t]
    \centering
    \includegraphics[width=0.8\columnwidth]{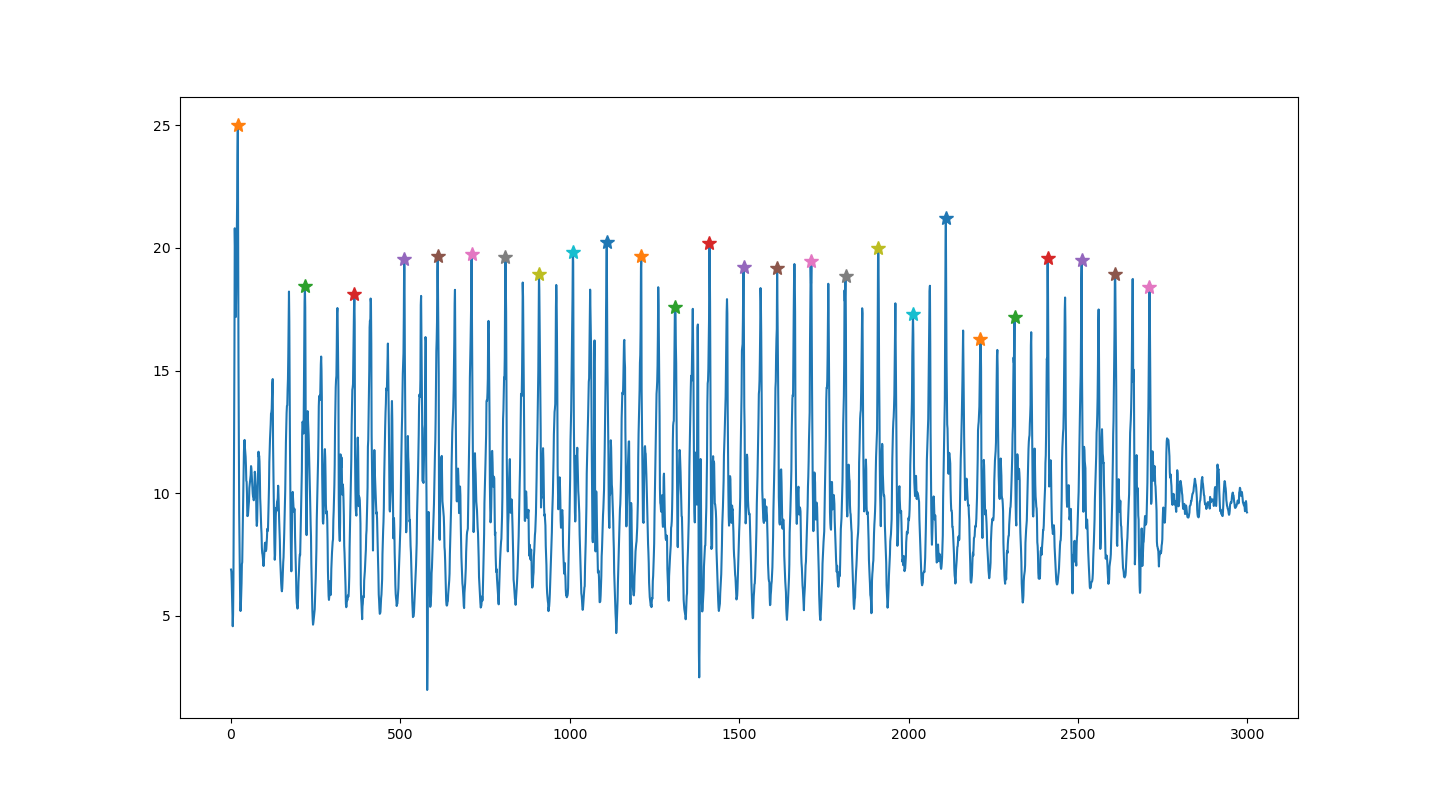}
    \caption{Mobile Phone Deployment: Segmentation}
    \label{fig:peaks}
\end{figure}

\begin{figure}[t]
    \centering
    \includegraphics[width=\columnwidth]{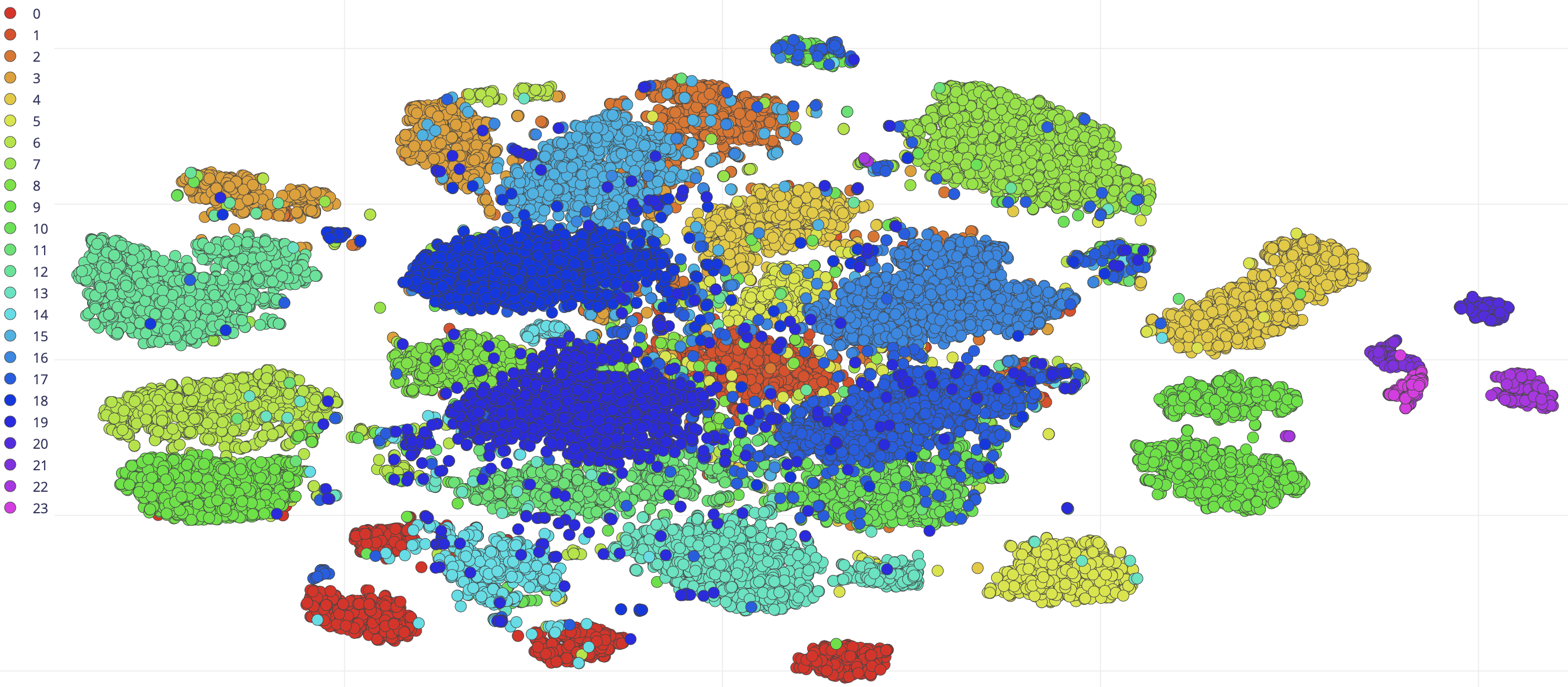}
    \caption{UMap Data Visualization}
    \label{datavis}
\end{figure}

\begin{figure}[t]
    \centering
    \includegraphics[width=0.7\columnwidth]{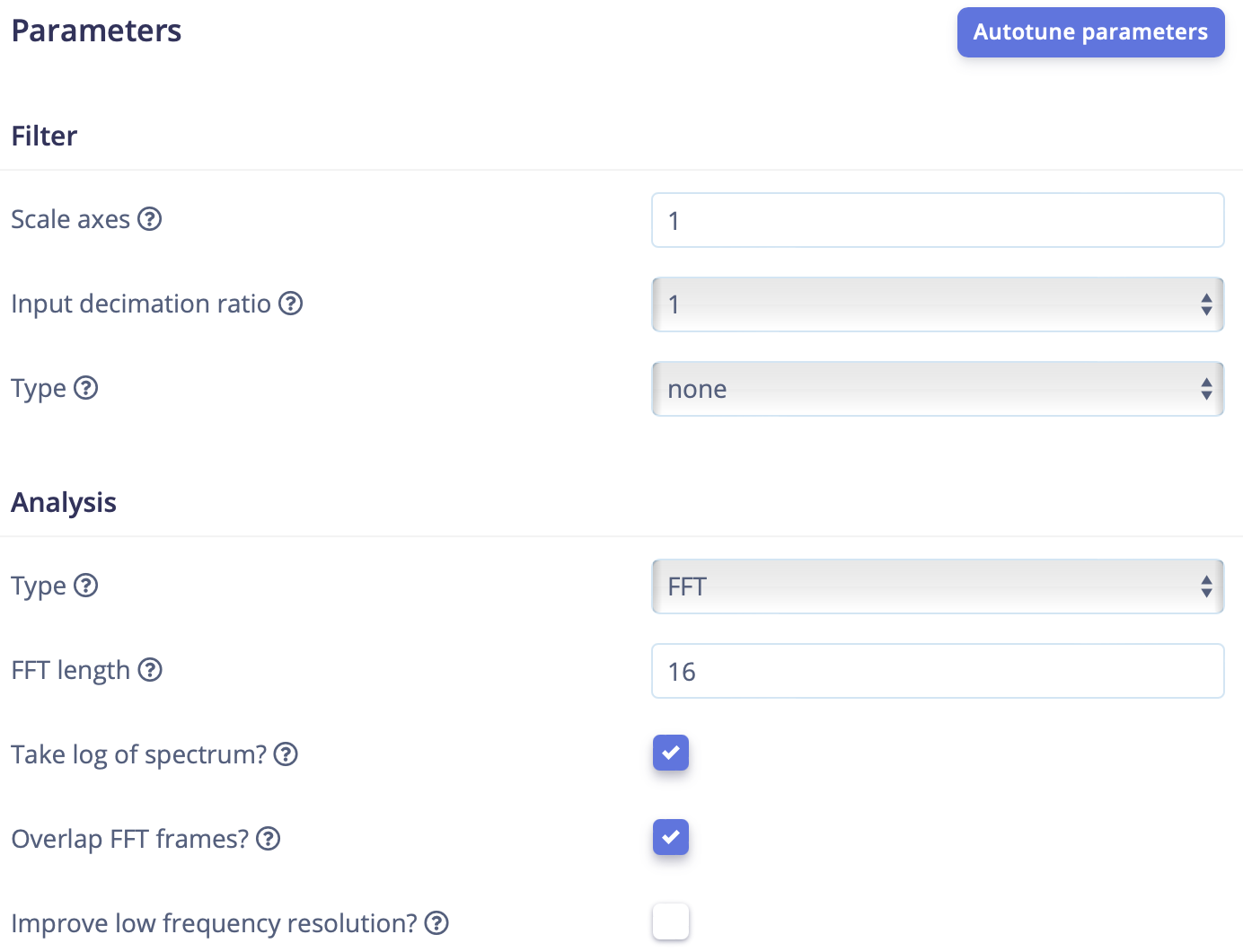}
    \caption{FFT Parameters}
    \label{fftparam}
\end{figure}

\begin{figure}[t]
    \centering
    \includegraphics[width=0.8\columnwidth]{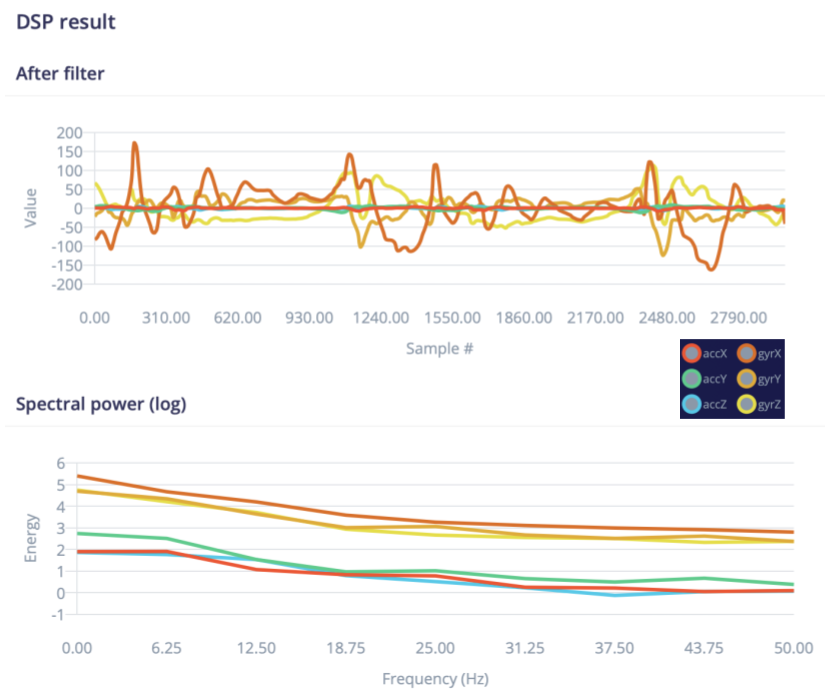}
    \caption{FFT Features}
    \label{fftfeat}
\end{figure}

To facilitate future expansion, we also implement an automated method for segmenting gait data that involves two steps: 1) employing a machine learning model to identify the walking period from all activities, including running, standing, and irregular movements, and 2) automatically splitting the gait segments based on data patterns. It is worth noting that we only perform such data processing during the training process to facilitate the collection of large amounts of data and improve the accuracy and stability of the model. During inference, we do not need to split the data since the window is sliding, and we can filter out non-walking data by setting a confidence threshold. Although including a data processing module to distinguish walking data in the inference process may improve the accuracy of the model, we determine it is not worth the computational cost.

We use a one-dimensional DCNN \cite{zou2020deep} model to roughly extract walking period based on its semantic difference with non-walking period. The result is shown in Figure \ref{fig:dcnnRes}. The blue zigzag line represents the collected activity data, which includes walking, stopping, and random movement in different directions and slopes. The green part of the straight line indicates a walking data segment, while the red part indicates a noise period that should be discarded. Afterwards, the extracted walking data was segmented into fixed-length segments to align them with other data in the training set. We exclude some atypical stride data based on peaks with fluctuating regularity and segment them into a series of 128-length two-step segments, as shown in Figure \ref{fig:peaks}.

\subsection{Inferencing Pipeline}

Our inferencing pipeline consists of two main components:
\subsubsection{Feature Extraction}
We use spectral analysis preprocessing block in Edge Impulse to extract spectral features from raw data samples. The corresponding UMap separation diagram visualizing the clusters for all 24 classes are shown in Figure \ref{datavis}. The FFT parameters used are shown in Figure \ref{fftparam}. We use FFT analysis with FFT length of 16 on a window size of 3 sec. Figure \ref{fftfeat} illustrates the result after filtering as well as logarithmic spectral power for an example sample. An interesting observation from Figure \ref{fftfeat} is that the gyroscope data carries more energy than the accelerometer data. This implies that collected gait data has richer features in angular momentum relative to acceleration. This observation is consistent for all data collected with Arduino. Future investigations can explore equalizing the energy between accelerometer and gyroscope for more uniform integration of signals.

\subsubsection{Model Architecture}
Our model architecture (Figure \ref{arch}) consists of 4 layers, where the first layer is a 2D Conv layer with 32 output channels and 3x3 filters. The input consists of a vector of 78 FFT-applied features which is reshaped to (13x6) before passing into the Conv layer. The Conv layer is followed by flatten, and subsequently 3 Dense layers with 256, 128 and 32 neurons respectively. The final layer is a softmax layer with 24 classes. The Conv layer extracts rich set of features from input, which are then mapped to the output classes through the 3 dense layers. This network was chosen after extensive hyperparameter tuning resulting in the best tradeoff between accuracy performance and model complexity.

\begin{figure}[t]
    \centering
    \includegraphics[width=0.75\columnwidth]{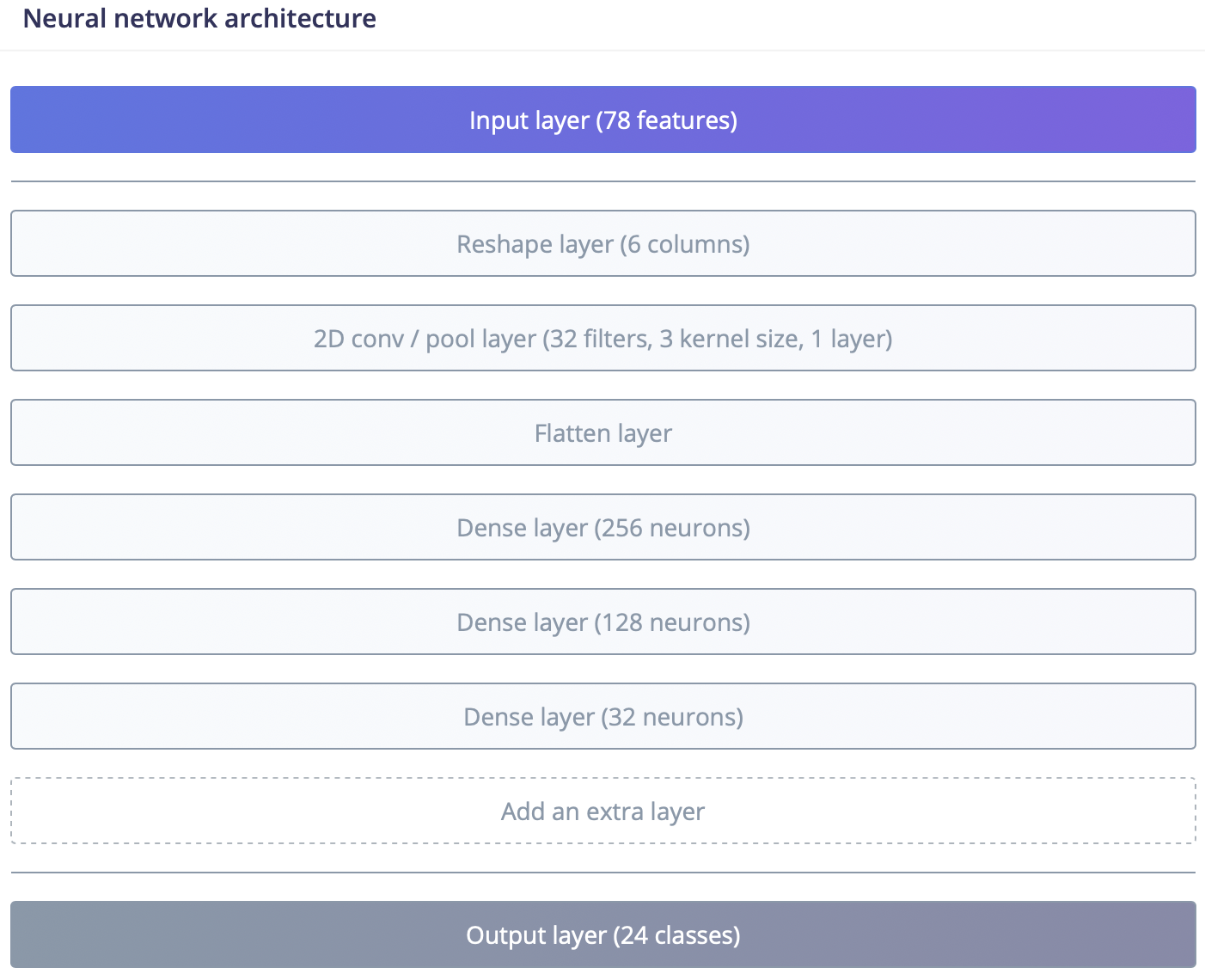}
    \caption{Neural Network Architecture}
    \label{arch}
\end{figure}

\begin{figure}[t]
    \centering
    \includegraphics[width=0.6\columnwidth]{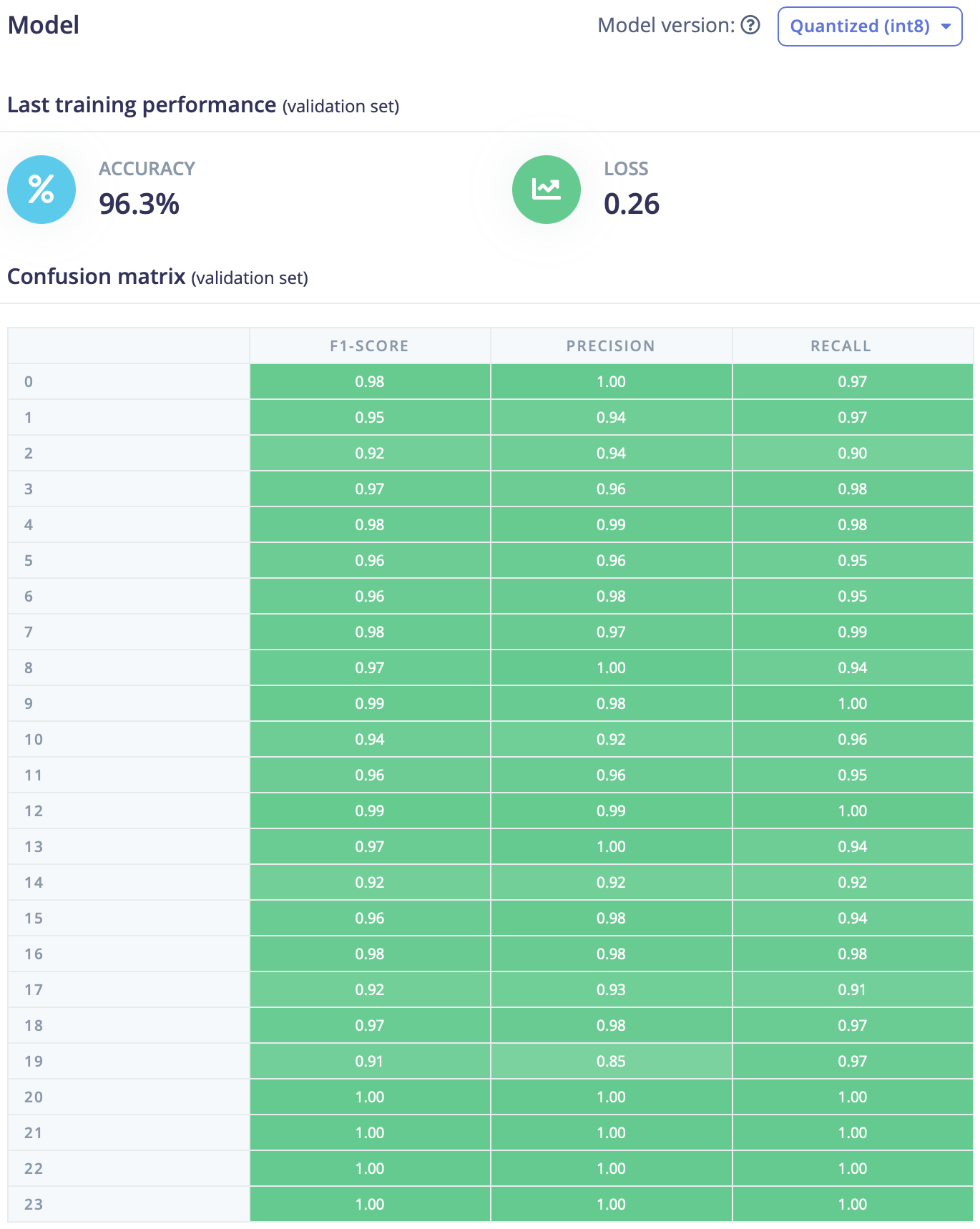}
    \caption{Validation Accuracy and Confusion Matrix}
    \label{val}
\end{figure}

\begin{figure}[t]
    \centering
    \includegraphics[width=0.6\columnwidth]{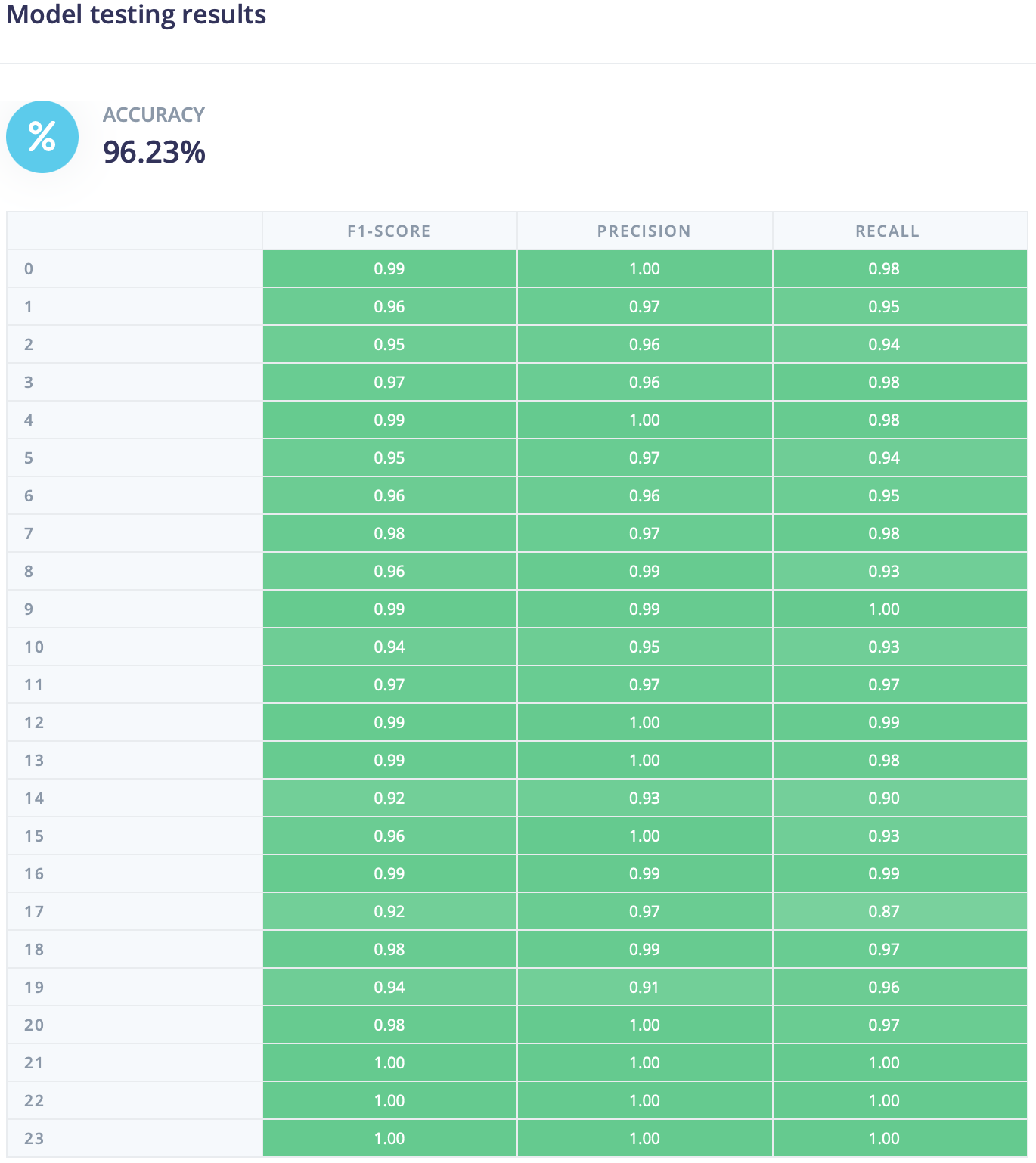}
    \caption{Testing Accuracy and Confusion Matrix}
    \label{test}
\end{figure}

\begin{figure}[t]
    \centering
    \includegraphics[width=0.8\columnwidth]{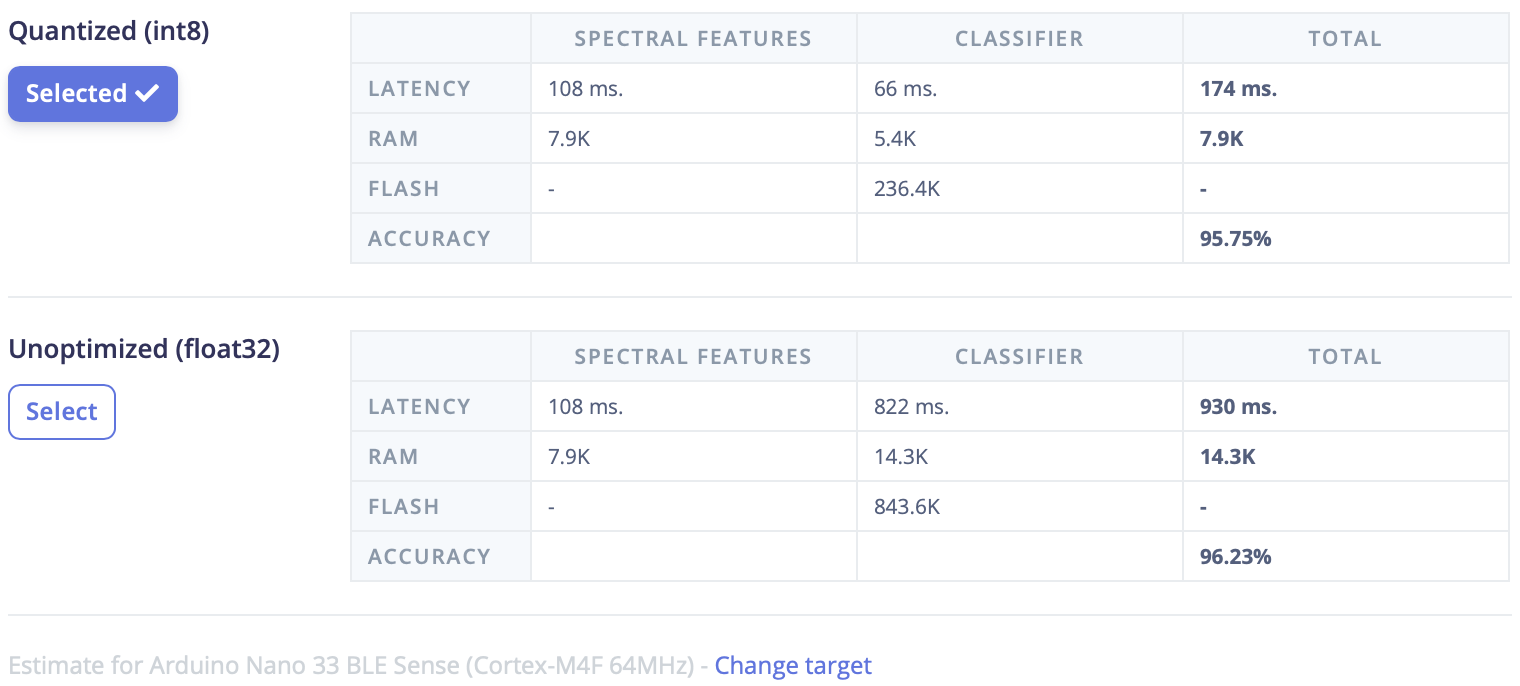}
    \caption{Deployment Metrics}
    \label{deploy}
\end{figure}


\subsection{Training}
\label{sec:training}

\subsubsection{Hyperparameters}
Sampling frequency for dataset collection is set to 100 Hz. FFT and window size parameters are described in previous sections. The model is trained for 20 epochs with 0.0005 learning rate. During training, 20\% of the training set is used for validation. Batch size is set to 32.

\subsubsection{Challenges and Iterations}
Two main challenges were faced during the course of this work. One was hyperparameter tuning such as window size etc. for manual data collection as well as manual window splitting for each of the new samples which is time-intensive. The second main challenge was associated with logistics and deployment code for live demonstration on Arduino. Parameters such as inferences per second had to be precisely tuned. During live demonstration, it was determined best for the subject to walk with the laptop in hand, wherein the laptop is connected to Arduino inside the pant pocket. Huge number of iterations had to be performed to arrive at the final data collection pipeline as well as the model architecture and training/deployment parameters.

\subsubsection{Results}
The trained model achieves 96.3\% validation accuracy and 96.23\% testing accuracy with almost perfect confusion matrix (Figures \ref{val} and \ref{test}). The INT8 deployment accuracy is 95.75\% with inference time of 66 ms, 5.4KB RAM usage, and 236.5KB Flash memory usage (Figures \ref{deploy}). The results underscore the strong feasibility of a lightweight model computationally powerful enough to classify 24 classes that can easily fit into a small form factor such as Arduino.

\section{Deployment Results}
We performed two types of live demonstration: 1) Arduino (main demo) and 2) smartphone. Further, we also deployed the model offline to BrainChip Akida.

\begin{figure}[t]
    \centering
    \includegraphics[width=0.8\columnwidth]{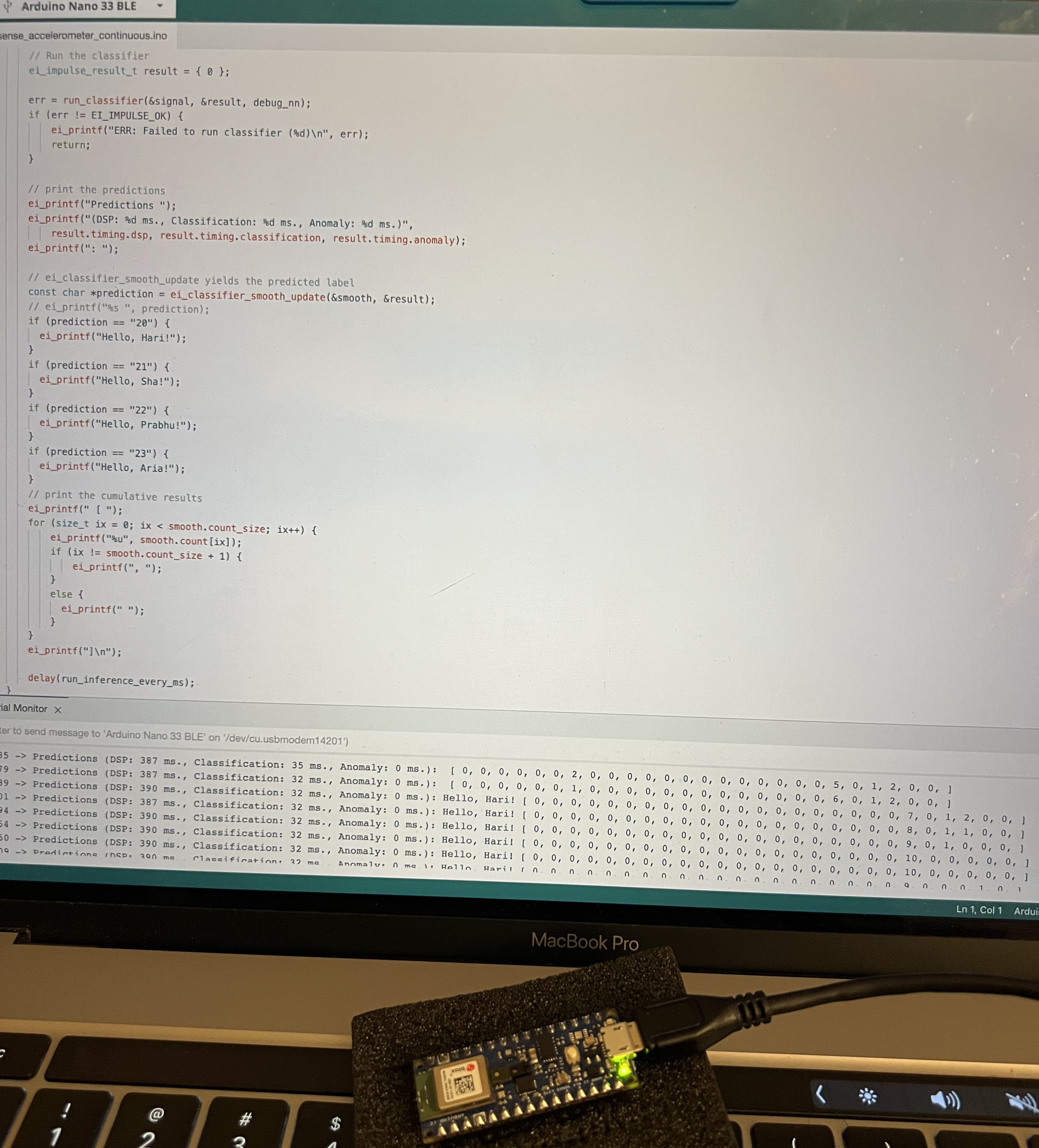}
    \caption{Arduino deployment code and output recognizing one team member when the Arduino was placed inside their pant pocket while walking.}
    \label{arduino_dep}
\end{figure}

\subsection{Deploy on Arduino Nano 33 BLE Sense}
During deployment on Arduino (Figure \ref{arduino_dep}), we successfully demonstrated accurate prediction of each of the four team members from live gait. We observed a 2 second delay from the onset of walking to the generation of appropriate predictions from the model. This overhead could be related to the prediction smoothing function within the Arduino deployment code and is a topic for future investigation and improvement. However, once it starts generating predictions, the inference latency is only about 70 ms with 5KB active RAM usage. Further, its power consumption was measured using a power jive to be 125 mW (25 mA current at 5 V).

\subsection{Deploy on Mobile Phone}
In order to simulate real-world application scenarios, we further experimented with deploying the model on a mobile device. Specifically, we deployed the model on an iPhone 13 using Edge Impulse. However, for this particular mobile device, the mobile motion sensor accessible by Edge Impulse only has three-axis accelerometer, and gyroscopic data cannot be retrieved. As a result, we adjusted the input features of the model from six to three in order to accommodate this limitation. Due to the differences in coordinate systems, precision, and amplification between the sensors on mobile devices and those on Arduino, the model deployed on the phone requires training with sensor data collected from the phone (following the same procedures as in the previous section). For demonstration, the inference model was designed to only recognize the gait of the mobile device owner. As shown in Figure \ref{fig:phoneDp}, label \textit{MyGait} denotes the gait of the phone owner; label \textit{0} represents a stationary pattern; labels \textit{1} and \textit{2} are others' gaits from the dataset. Once the mobile device owner starts walking steadily, the model is capable of identifying the identity based on the gait.

\begin{figure}[t]
    \centering
    \includegraphics[width=0.65\columnwidth]{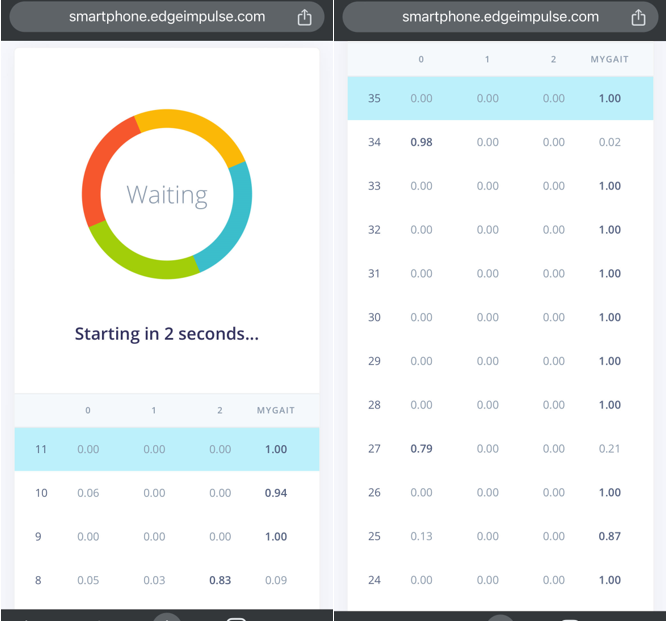}
    \caption{Mobile Phone Deployment}
    \label{fig:phoneDp}
\end{figure}

\subsection{Deploy on BrainChip Akida}
As a third alternative, we convert the trained CNN to an equivalent SNN using BrainChip MetaTF and assess inference metrics on remotely accessible BrainChip Akida processors (physically located in CMU's Silicon Valley campus). As shown in Figure \ref{akida}, the converted SNN mapped to BrainChip Akida processor consumes about 880 mW with an average framerate of 22.73 fps. The inference energy consumed is 45.92 mJ/frame. This is only a preliminary analysis that needs deeper investigation for power and energy optimization.

\begin{figure}[t]
    \centering
    \includegraphics[width=\columnwidth]{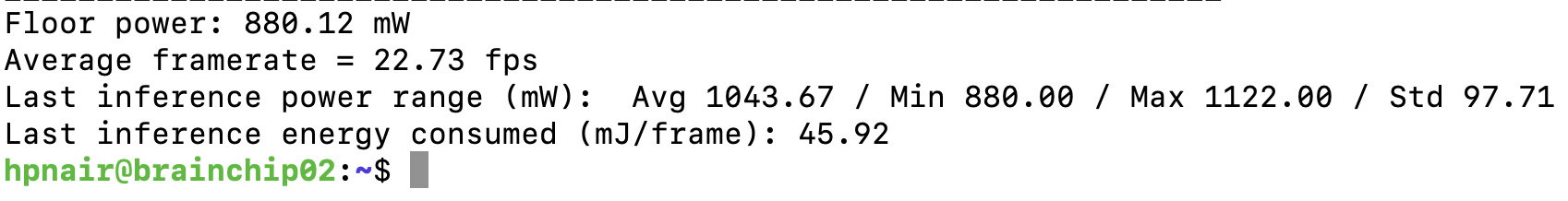}
    \caption{BrainChip Akida Inference Metrics}
    \label{akida}
\end{figure}

\section{Discussion}
 
\subsection{BLERP Model}
BLERP offers a multi-dimensional perspective to assess the advantages and trade-offs of edge-based solutions in various application domains, as explained below.

\textbf{Bandwidth:} In gait analysis, it is crucial to collect and analyze large amounts of data to track and monitor the movements of the joints accurately. However, communicating all this data to a cloud server can be bandwidth-intensive. Edge-based models can operate on local data, reducing the amount of data transmission and alleviating bandwidth pressure.

\textbf{Latency:} Since gait recognition is used to authenticate the user, real-time sensing is important, and transmitting data to and from the cloud would incur additional latency. Processing the data locally on the device mitigates this issue, allowing the solution to collect data and perform analysis in real time.

\textbf{Economics:} Gait analysis can leverage inexpensive IMU sensors, pre-existing in most wearables. TinyML serves as a cost-effective solution, as it allows for the deployment of machine learning models on low-cost, low-power devices.


\textbf{Reliability:} Human gait has been demonstrated to be difficult to impersonate, thereby making gait recognition relatively robust to biometric duplication and more reliable than other biometric approaches in terms of security. Further, edge processing enables the solution to reliably deliver results, even in areas with poor network connectivity.

\textbf{Privacy:} Wearable IMU sensors can be used to collect data without invading user privacy and operate on the edge without sending sensitive data to cloud, ensuring data privacy.

\subsection{Ethical Challenges}
\textbf{Privacy:} Although processing data locally on a device can help protect the privacy of bioinformation, it is still important to ensure that the data collected is not used to invade user privacy. This includes taking measures to ensure that the data cannot be traced back to a specific individual or used for purposes other than real-time gait analysis. 

\textbf{Informed Consent:} When using wearable IMU sensors to collect gait data, obtaining informed consent from the participants is important. This includes providing clear information about the purpose of the data collection, how the data will be used, and who will have access to the data.

\textbf{Bias:} Bias can occur when the dataset used to train the algorithm does not represent the population it intends to serve or contains unintentional bias. This can lead to inaccurate or unfair results for certain groups, particularly  individuals with different body types or physical disabilities . To mitigate these challenges, it is important to prioritize diverse and representative datasets, actively collecting data from individuals of different ages, ethnicities, body types, and abilities.

\section{Conclusion}
Our work serves as a feasibility proof for deployment of very efficient yet highly effective lightweight models on to small form factor edge devices such as Arduino, smartphone, etc. to identify person using their gait. Our model trained on a standard dataset augmented with the team members' gait data is able to achieve  96\% accuracy on 24 classes, while consuming only 70 ms inferencing time, 5 KB RAM, and 125 mW power. Further, our work serves as a first step towards deploying a gait recognition model on a neuromorphic device such as BrainChip Akida. Future investigation will focus heavily on optimizing the model, adding data for physically challenged as well as multiple other people for diversification and bias reduction, and optimizing the neuromorphic model.

\bibliographystyle{IEEEtran}
\bibliography{refs}

\end{document}